# Geometric origin of probabilistic distributions in statistical mechanics

C. Vignat† and A. Plastino‡

† L.I.S. Grenoble, France and E.E.C.S., University of Michigan, USA, e-mail: vignat@univ-mlv.fr
‡ La Plata National University & National Research Council (CONICET)
C. C. 727 - 1900 La Plata - Argentina, e-mail: plastino@fisica.unlp.edu.ar

**Abstract.** A finite dimensional-system whose physics is governed by a Gaussian distribution can be regarded as a *subsystem* of an infinite dimensional-underlying system described by a *uniform* distribution on the (infinite dimensional) sphere. In turn, a finite dimensional-system ruled by a power-law probability distribution can be considered as a *subsystem* of a finite underlying system whose physics is governed by a uniform distribution on a finite dimensional sphere. These two uniform distributions (on finite and infinite dimensional spheres) are the most "natural" distributions since they maximize both Tsallis and Shannon entropies without any extra constraint but bounded support. The main advantage of power-law distributions could thus be assigned to the fact that they correspond to a still more "natural" underlying model, namely a finite dimensional sphere. In many physical situations, the underlying system is so large that the distinction between finite and infinite dimension is difficult to see, so that the Boltzmann-Gibbs distribution acquires a great degree of versatility. Moreover, an explicit value of the power-law exponent can be given in terms of the dimension of the underlying model and a geometric interpretation of the temperature-concept is provided.

## 1. Introduction

Statistical mechanics arises as the physicist' answer to the problem of solving dynamical problems without enough available information (for example, about initial conditions) [1, 2]. Probability distributions (PDs) become the preferred mathematical tool for the purpose of system's description. In turn, these PDs are obtained according to several procedures like ensemble construction, steepest-descent, or Maximum Entropy (MaxEnt) treatments [1, 2, 3, 4]. Here we wish to provide a geometric interpretation to these all-important PDs.

## 2. A result by Poincaré

Recall that the sphere of radius $n^{1/2}$ in $R^n$ writes $\mathcal{S}_n(n^{1/2}) = \{x \in R^n | \sum_{i=1}^n x_i^2 = n\}$. The following important result is due to Poincaré [5], and a simple proof of it can be found in [6]:

If the vector $X = (X(1), \ldots, X(n))$ is UNIFORMLY distributed on the sphere of radius $n^{1/2}$ in $R^n$, then, for fixed $m < \infty$, the probability $P(X(1), \ldots, X(m))$ of encountering $X(i)$ located in the interval $[a(i), b(i)]$ $(1 \leq i \leq m)$ is, in the limit $n \to \infty$, given by

$$P(X(1), \ldots, X(m)) = \prod_{i=1}^m \int_{a(i)}^{b(i)} dy \, \frac{\exp(-y^2/2)}{\sqrt{2\pi}}. \tag{1}$$

We wish here to give the following *interpretation* to the above result: suppose that our vector $X$ describes some physical property of an $m-$ dimensional system and that $X$ is Gaussian-distributed. We are authorized by Poincaré to regard our system as a *sub*-system of an infinite system uniformly distributed on the sphere in $R^\infty$. Such an interpretation allows for what we believe to be a new and elegant derivation of Gibbs' canonical ensemble, as advanced in what follows.

### 2.1. Gibbs' canonical ensemble constructed à la Poincaré

Consider now a scenario in which our system is described by a general quadratic hamiltonian form for a system with $m$ degrees of freedom ($m/2$ "coordinates" $x$ and $m/2$ "momenta" $p$)

$$H = \frac{1}{2} \sum_{i,j=1}^{m/2} T_{ij} \, p_i p_j + V_{ij} \, x_i x_j + W_{ij} \, p_i x_j + W_{ij}^t \, x_i p_j, \tag{2}$$

where $W$, $T$ and $V$ are $(m/2 \times m/2)-$matrices, the last two symmetric ones, and the supra-index $t$ stands for transposed ($x_i$ and $p_i$ are, of course, classical variables). We envision now, following Gibbs' ideas [1], $(N-1)$ identical copies of the system described by the Hamiltonian (2), with $N = n/m$ and $n \to \infty$.

We have thereby constructed our Gibbs' ensemble of $N$ identical systems. Since at this stage we know nothing about probabilities for the systems in the ensemble, we are forced to assign to it a uniform distribution according to Bernoulli's principle of insufficient reason [3, 4].

In the above considered vector $X = (X(1), \ldots, X(N))$, each $X(i)$ becomes partitioned in the fashion

$$X(i) \equiv X(i)_1, \ldots, X(i)_m; \quad (i = 1, \ldots, N). \tag{3}$$

After passing to an $m-$normal modes representation via Hamiltonian-diagonalization, i.e., $H = \sum v_j^2$ (for each of the $N$ copies labelled by $i$), we immediately appreciate the fact that the Gaussian probability distribution of Eq. (1) yields the celebrated Gibbs' canonical one [1, 2], with the identification $X(i)_j \equiv v_j/kT$, for all $i = 1, \ldots, N$. The constant $k$ is Boltzmann's one while $T$ is the temperature.

In terms of Information Theory [3, 4] tenets, we face then this situation:

- without MaxEnt, we have a system uniformly distributed on the sphere in $R^\infty$. Accordingly [4], our ignorance is maximal. With the sole information that i) the system has $m$ degrees of freedom, ii) the Hamiltonian $H$ is quadratic, and iii) the temperature is $T$, we can ascertain that our scenario is ruled by the BG canonical distribution. Our ignorance has thereby been reduced (information gained) to that provided by the Shannon entropy associated to the BG canonical distribution. Notice that $T$ has to be given from the "outside".
- With MaxEnt. Same as above, but item iii) is replaced by knowledge of $\langle H \rangle$. The variational procedure yields then $T$. Nothing from the "outside" is needed.

Instead of considering an infinitely large heat reservoir, as Gibbs does, Poincaré' theorem provides us with the infinite system of which *ours* is a subsystem.

The above considerations are not restricted to a classical treatment. Consider the general quantal quadratic hamiltonian form for a system with $m-$degrees of freedom

$$\hat{H} = \frac{1}{2} \sum_{i,j=1}^n T_{ij}\hat{p}_i\hat{p}_j + V_{ij}\hat{x}_i\hat{x}_j + W_{ij}\hat{p}_i\hat{x}_j + W_{ij}^t\hat{x}_i\hat{p}_j, \tag{4}$$

where, as above, $W$, $T$ and $V$ are $n \times n-$matrices, the last two symmetric ones. The preceding discussion applies in its totality to the quantal scenario.

## 3. Power law probability distributions and general hamiltonian forms

The preceding considerations have been restricted to quadratic hamiltonians. This restriction can be lifted, by recourse to arguments invoking power-law distributions, to more general hamiltonian forms that, after a suitable diagonalization process, end-up with a Hamiltonian consisting of $k$ terms,

$$H_{diagonal} = \sum_{i=1}^k v_i. \tag{5}$$

Power-law distributions are ubiquitous in physics, critical phenomena being a conspicuous example [7]. In a statistical mechanics' context they arise quite naturally if the information measure one maximizes (subject to appropriate constraints) in order to arrive at the equilibrium distribution is not Shannon's one but a generalized one [8, 9].

*3.1. Two important theorems*

We say that a random vector $Z$ is orthogonally invariant if, for any deterministic orthogonal matrix $A$, vector $AZ$ is distributed as $Z$. A typical physical example is that of $R^3-$rotations, that are represented by orthogonal matrices. Obviously, the physical meaning attached to the $Z-$distribution will not change if we rotate the coordinate-system [10].

A vector uniformly distributed on the unit sphere in $R^n$ can be obtained by normalizing a vector distributed according to *any* orthogonally invariant distribution.

**Proposition 1** *An n-variate random vector $U$ is uniformly distributed on the unit sphere in $R^n$ if it writes*

$$U = \frac{Z}{\|Z\|_2}$$

*where $Z$ is orthogonally invariant [11].*

As it is clear from the above, vector $U$ has unit 2-norm:

$$\|U\|_2 = \left[\sum_{i=1}^n |u_i|^2\right]^{1/2} = 1,$$

which is to be regarded as a constraint in a MaxEnt variational procedure.

*3.1.1. the case $q > 1$ or $q < 0$* The marginal distributions of a uniform distribution on the unit sphere in $R^n$ *are power-law ones* with $q > 1$ or $q < 0$ and can be easily computed, as specified by the following theorem.

**Theorem 2** *if $U$ is uniformly distributed on the unit sphere in $R^n$ then the marginal density of $V = [U_1, \ldots, U_k]^T$ is the power law density*

$$f_V(u_1, \ldots, u_k) = \frac{p^k \Gamma\left(\frac{n}{2}\right)}{2^k \Gamma^k\left(\frac{1}{2}\right) \Gamma\left(\frac{n-k}{2}\right)} \left(1 - \sum_{i=1}^k |u_i|^2\right)^\tau$$

$$\text{with } \tau = \frac{n-k}{2} - 1 \text{ and } 1 \leq k \leq n-1. \tag{6}$$

The proof of theorem 2 can be found in [12]. The distribution (6) is often called nowadays a Tsallis' one [8, 9], characterized by the so-called nonextensivity parameter $q$ [8, 9] that in the present instance is given by

$$q = \frac{n-k}{n-k-2}. \tag{7}$$

Tsallis' density distributions maximize, with appropriate constraints, the entropy or information measure (a functional of the distribution $f(X) \equiv f(x_1, \ldots, x_k)$)

$$S_q[f] = \frac{1}{q-1}\{1 - \int dX\,[f(X)]^q\}. \tag{8}$$

Indeed, the probability distributions corresponding to Th.2 maximize Tsallis' entropy under a constraint on their order-2 moment, i.e., given $k \leq n-1$ real, fixed numbers $K_i$ as constraints:

$$\int X_i^2 f(X) dX = K_i \tag{9}$$

the maximizer of $S_q[f]$ writes (see [14])

$$f(x_1, \ldots, x_k) = (1 - \sum_{i=1}^{k} \lambda_i x_i^2)^{(n-k)/2-1} \tag{10}$$

where $\lambda_i$ are the $k$ positive associated Lagrange multipliers. This amounts to stretching any component $u_i$ by a factor $(\lambda_i)^{1/2}$. Moreover, if $[u_1, ..., u_n]$ is uniform *on* the unit sphere in $R^n$, then

ALL ITS MARGINALS MAXIMIZE TSALLIS' ENTROPY WITH A POSITIVE PARAMETER $q$ PROVIDED THAT $(n-k)/2 - 1 > 0$ OR, EQUIVALENTLY, THAT $1 \leq k \leq n-2$.

*3.1.2. the case $n/(n+2) < q < 1$* We first recall a general result about maximizers of Tsallis entropy under order-2 moment constraint in the case $n/(n+2) < q < 1$. Its proof can be found in [14].

**Proposition 3** *Given n constraints*

$$\int Y_i^2 f(Y) dY = K_i \tag{11}$$

*the maximizer of $S_q[f]$ with $n/(n+2) < q < 1$ writes*

$$f_Y(y_1, \ldots, y_n) = (1 + \sum_{i=1}^{n} \lambda_i y_i^2)^{-(m+n)/2} \tag{12}$$

*where $\lambda_i$ are the n positive associated Lagrange multipliers and parameter m is defined as*

$$m = \frac{2}{1-q} - n \tag{13}$$

*If moreover parameter $m \in N$ then a stochastic representation of random vector Y writes as follows*

$$Y^t = \frac{[N_1, \ldots, N_n]}{\sqrt{M_1^2 + \ldots + M_m^2}} \tag{14}$$

*where $\{M_i\}$ and $\{N_i\}$ are independent unit variance random variables.*

Power-law distributions with $n/(n+2) < q < 1$ can be obtained from a uniform distribution on the unit sphere in $R^n$ by using the following duality result:

**Proposition 4** *If X is a k-variate marginal of a vector U uniformly distributed on the unit sphere in $R^n$ ($n > 2$) then the k-variate vector Y defined as*

$$Y = \frac{X}{\sqrt{1 - X^t X}} \tag{15}$$

*follows a Tsallis distribution with parameter $q = \frac{n-2}{n}$*

**Proof.** A stochastic representation for vector $X$ writes

$$X^t = \frac{[N_1, ..., N_k]}{\sqrt{N_1^2 + \ldots + N_n^2}} \tag{16}$$

where $N_i$ are independent Gaussian random variables with unit variance, so that vector $Y$ defined by (15) has the following stochastic representation

$$Y^t = \frac{[N_1, ..., N_k]}{\sqrt{N_{k+1}^2 + \ldots + N_n^2}} \tag{17}$$

Thus it maximizes Tsallis entropy with parameter $q$ such that

$$\frac{1}{q-1} = -\frac{k+(n-k)}{2} \Leftrightarrow q = \frac{n-2}{n} \tag{18}$$

∎

Recall that the the unit ball $\mathcal{B}_k$ writes $\mathcal{B}_k = \{x \in R^k | \sum_{i=1}^k x_i^2 \leq 1\}$. The duality relationship as expressed by (15) allows us to derive a geometrical construction of Tsallis distributed random vectors in the case $n/(n+2) \leq q \leq 1$ as follows.

**Theorem 5** *if vector $X = OM$ belongs to the unit ball $\mathcal{B}_k$, then point $M$ can be considered as the orthogonal projection on $\mathcal{B}_k$ of point $P \in \mathcal{S}_{k+1}(1)$. The intersection of line $[OM]$ with the hyperplane $\mathcal{H}_{k+1} = \{Z \in R^{k+1} | Z_{k+1} = 1\}$ defines a unique point $N$ such that vector $Y = [(ON)_1, \ldots, (ON)_k]^t$ verifies relation (15), and thus follows a Tsallis distribution with parameter $q = (n-2)/n$.*

**Proof.** A usual parameterization of the sphere $S_{k+1}(1)$ writes $\forall j \in \{1, \ldots, k+1\}$

$$X_j = \cos\theta_{j-1} \prod_{i=j}^k \sin\theta_i \tag{19}$$

with the convention $\theta_0 = 0$ and $\prod_{i=k+1}^k = 1$. As $X \in \mathcal{B}_k$, $\exists \theta_k$ such that $\sum_{i=1}^k X_i^2 = \sin^2\theta_k$, and defining $X_{k+1} = \cos\theta_k$ we deduce that $X$ is the orthogonal projection on $\mathcal{B}_k$ of $OP = [X_1, \ldots, X_{k+1}]^t \in \mathcal{S}_{k+1}(1)$. Moreover, $\exists \alpha \in R$ such that $ON = \alpha OP$ but as $(ON)_{k+1} = 1$ then necessarily

$$(ON)_i = \frac{X_i}{X_{k+1}} \forall i \in \{1, \ldots k\} \tag{20}$$

so that finally

$$Y^t = \frac{[X_1, \ldots, X_k]}{X_{k+1}} = \frac{X^t}{\sqrt{1 - X^t X}} \tag{21}$$

∎

We remark that, by symmetry, hyperplane $\mathcal{H}_{k+1}$ can be replaced by any hyperplane $\mathcal{H}_i$ with $1 \leq i \leq k$. This construction is illustrated on Fig.1 in the case $k = 2$: in this particular case, point $N$ is known in the field of geometry as the *gnomonic* projection of point $P$.

*3.2. Gibbs-Poincaré construction*

We devise the ensemble as in the preceding section with just one difference: the number of copies is now large but definitely *finite*: vector $X$ of Proposition 1 has $n$ components. These components are the $k$ terms of the pertinent Hamiltonian (Cf. Eq. (5)), *conveniently repeated*, as explained in (3).

*3.3. Fixing the value of the nonextensive index $q > 1$ or $q < 0$*

Geometric considerations suffice to fix $q$, a problem that has eluded solution until now [8, 9]! We see that only those marginals of dimensions $1, 2, ..., n-3$ are Tsallis maximizers with a finite positive parameter $q$. Additionally, the marginal of dimension $n - 2$ is uniform *in* (not *on*!) the unit sphere $\mathcal{S}_{n-2}(1)$, that is, $u_1^2 + ... + u_{n-2}^2 \leq 1$ (not $= 1$) and thus maximizes Tsallis' entropy, but with $q = +\infty$.

The marginal distribution for which $k = n - 1$ maximizes Tsallis entropy with a negative parameter $q = -\frac{1}{2}$. Summing up: as the dimension of the marginal decreases, we go from maximizers of Tsallis entropies with

- (A) $q < 0$ if $k = n - 1$
- (B) $q = +\infty$ if $k = n - 2$
- (C) $q > 1$ (finite) if $k \leq n - 3$
- (D) $q \simeq 1$ for $n \to \infty$.

For macroscopic systems, item (D) applies (orthodox Boltzmann-Gibbs statistics). In this limit the power-law distribution becomes the usual exponential one [8, 9] that we have discussed in Section II. Consider then the $n-$finite instance.

In most cases of physical relevance, *$k$ is small*, $k \ll n$, so that item (C) applies. Item (A) corresponds to a situation in which we have a great deal of information (because $n \gg 1$ and we know the distribution function for $n$ dimensions to be uniform), that specifies the most important aspects of the problem. Only small details remain to be determined. A distribution with $q < 0$, precisely, amplifies those small details [8, 9]. Item (B) seems to correspond to very peculiar situations. However, the $q = \infty$ distributions have found some uses in the theory of quantum entanglement [13].

*3.4. Geometrical character of the temperature*

As previously stated, a scenario akin to the Boltzmann-Gibbs one of the preceding section arises if we consider the Hamiltonian (2) (or (4)) expressed as a sum of $k$ quadratic terms (normal modes) and just a single MaxEnt constraint, namely, $\langle H \rangle$, with an associate Lagrange multiplier $\beta$ (inverse temperature). We have to deal now with Tsallis' distributions for the canonical ensemble that correspond to a more "natural" underlying model than the Gaussian distributions, since they are marginals of a uniform distribution on a *finite dimensional sphere* (Gaussians being marginals of unit distribution on an infinite dimensional one).

We appreciate the fact that the temperature $T$ acquires now a geometric interpretation. The square root of its inverse, namely, $\sqrt{\beta}$, is a stretching factor (Cf. Eq. (10) and the comments that follow). All $m$ components of vector $U$ in (10) are stretched by this factor. We can imagine that temperature arises as a result of the deformation induced in these components by the need to accommodate the mean energy constraint.

## 4. Conclusions

In this work we have provided a geometric interpretation of the canonical ensemble on the basis of Poincaré's ideas. Extension of this interpretation to more general ensembles (grand-canonical, etc.) is straightforward. Summing up, we have

(i) devised the Gibbs-Poincaré ensemble construction methodology,
(ii) given a geometric interpretation to the temperature,
(iii) established that a finite dimensional-system whose physics is governed by a Gaussian distribution can be regarded as a *subsystem* of an infinite dimensional-underlying system described by a *uniform* distribution on the (infinite dimensional) sphere,
(iv) in turn, a finite dimensional-system ruled by a power-law probability distribution can be considered as a *subsystem* of a finite underlying system whose physics is governed by a uniform distribution on a finite dimensional sphere,
(v) ascertained that these two uniform distributions (on finite and infinite dimensional spheres) maximize both Tsallis and Shannon entropies, and
(vi) given an explicit value to the power-law exponent in terms of the dimension of the underlying model.

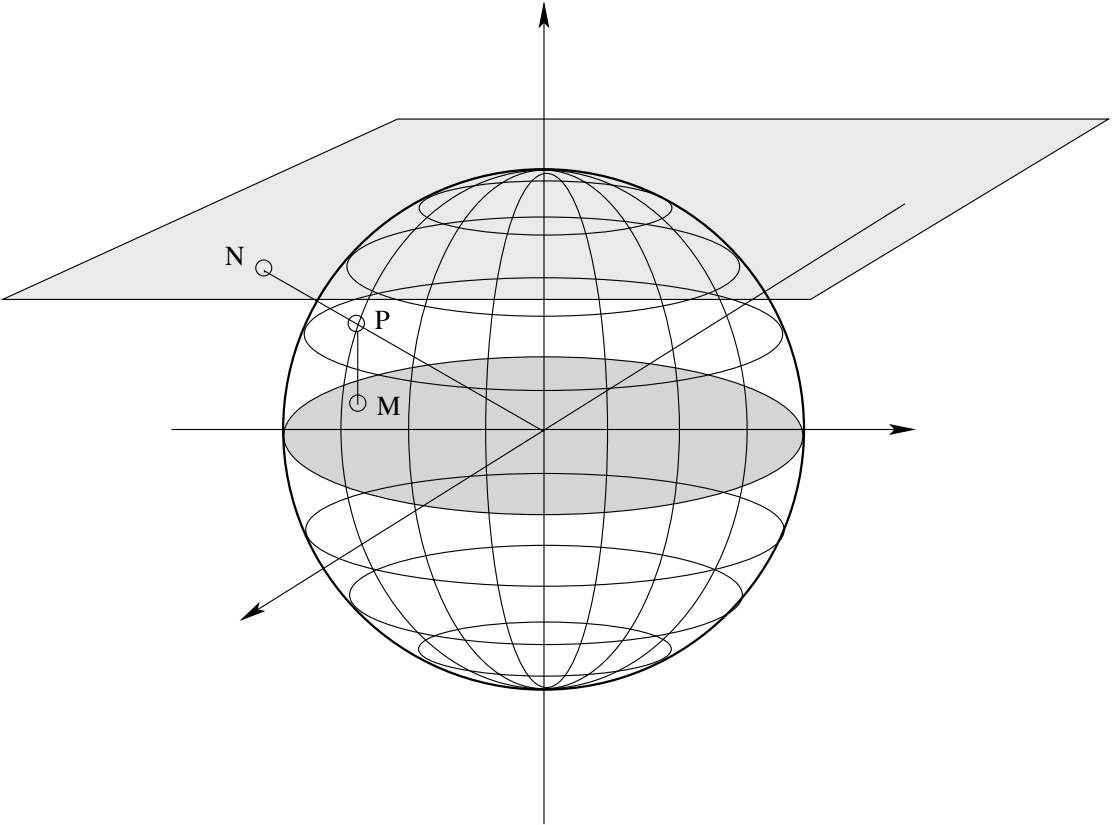

**Figure 1.** If vector OM is distributed according to the marginal of a uniform vector on the sphere in $R^n$, vector ON follows a Tsallis distribution with parameter $q = (n-2)/n$.